\documentclass[aps,prl,showpacs,twocolumn,superscriptaddress]{revtex4}
\usepackage{bm,color}
\usepackage{graphicx}
\usepackage{amsmath}
\pdfoutput=1

\begin{document}
\title {Theory of Magnetic-Field-Induced Polarization Flop in Spin-Spiral Multiferroics}

\author{Masahito Mochizuki}
\affiliation{$^1$ Department of Physics and Mathematics, Aoyama Gakuin University, Sagamihara, Kanagawa 229-8558, Japan}
\affiliation{$^2$ PRESTO, Japan Science and Technology Agency, Kawaguchi, Saitama 332-0012, Japan}
\begin{abstract}
The magnetic-field-induced 90$^{\circ}$ flop of ferroelectric polarization $\bm P$ in a spin-spiral multiferroic material TbMnO$_3$ is theoretically studied based on a microscopic spin model. We find that the direction of the $\bm P$-flop or the choice of $+\bm P_a$ or $-\bm P_a$ after the flop is governed by magnetic torques produced by the applied magnetic field $\bm H$ acting on the Mn spins and thus is selected in a deterministic way, in contradistinction to the naively anticipated probabilistic flop. This mechanism resolves a puzzle of the previously reported memory effect in the $\bm P$ direction depending on the history of the magnetic-field sweep, and enables controlled switching of multiferroic domains by externally applied magnetic fields. Our Monte-Carlo analysis also uncovers that the magnetic structure in the $\bm P$$\parallel$$\bm a$ phase under $\bm H$$\parallel$$\bm b$ is not a so-far anticipated simple $ab$-plane spin cycloid but a conical spin structure.
\end{abstract}
\maketitle

\section{Introduction}
\begin{figure}[t]
\includegraphics[width=1.0\columnwidth]{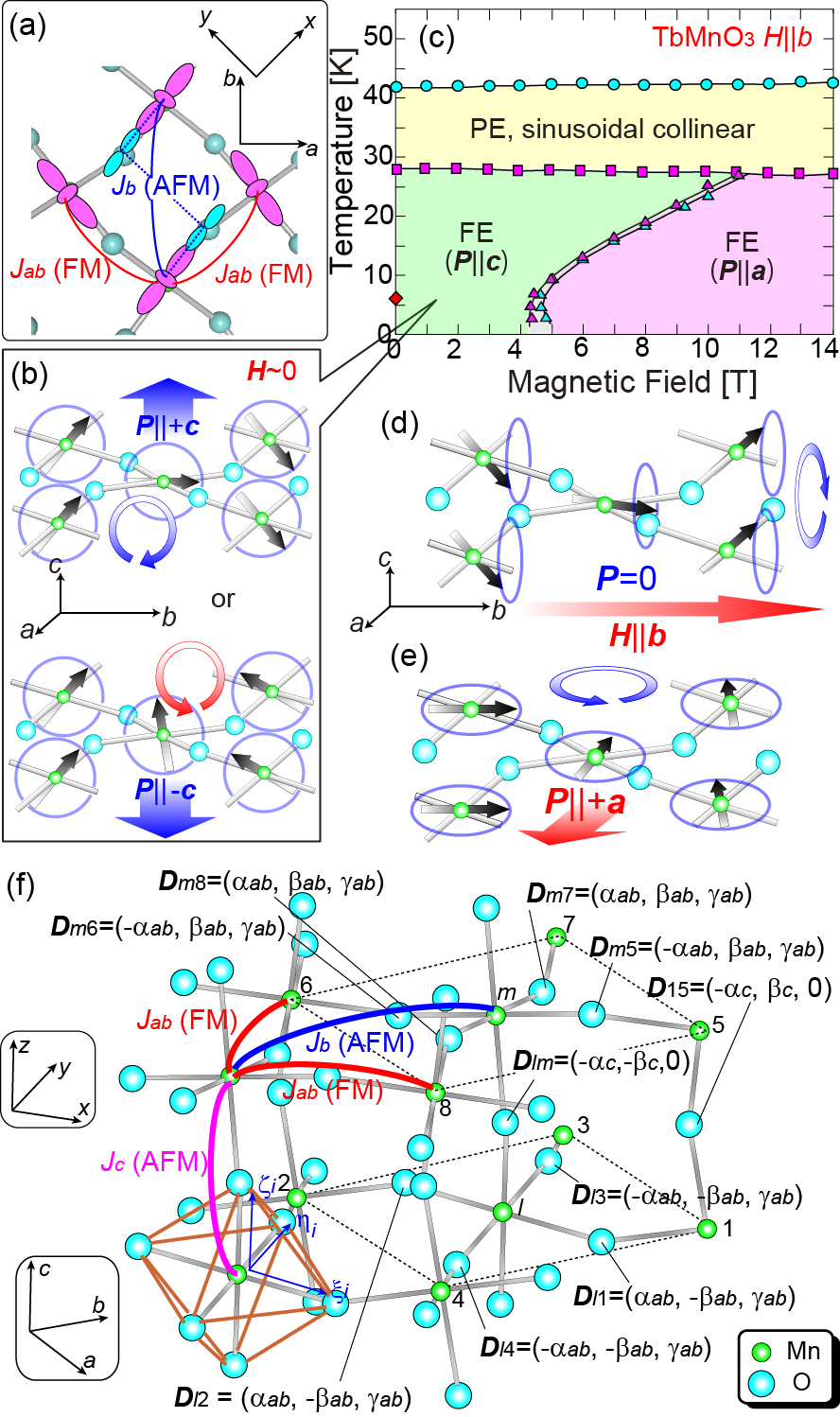}
\caption{(color online). (a) Frustrated ferromagnetic (FM) and antiferromagnetic (AFM) interactions in the $ab$ plane resulting from a staggered orbital ordering. (b) Clockwise and counterclockwise $bc$-plane cycloidal Mn-spin orders with $+\bm P_c$ and $-\bm P_c$. (c) Phase diagram of TbMnO$_3$ for $\bm H$$\parallel$$\bm b$. (d) Longitudinal conical spin state with $\bm P$=0 propagating along the $b$ axis. (e) $ab$-plane cycloidal spin order with $\bm P$$\parallel$$\bm a$. (f) Exchange interactions, Moriya vectors, and local axes ($\xi_i$, $\eta_i$, $\zeta_i$) attached to the tilted $i$th MnO$_6$ octahedron. For the exchange interactions, inter-plane antiferromagnetic exchanges $J_c$ as well as the frustrated $J_{ab}$ and $J_b$ in the $ab$ plane are considered. The Moriya vectors associated with the Mn-O-Mn bonds are expressed using five parameters, $\alpha_{ab}$, $\beta_{ab}$, $\gamma_{ab}$, $\alpha_c$ and $\beta_c$.}
\label{Fig1}
\end{figure}
Nontrivial spin textures in magnets often host intriguing physical phenomena through couplings to the charge degrees of freedom~\cite{Nagaosa12}. Multiferroics, i.e., magnetism-induced ferroelectricity, is a typical example~\cite{Kimura03a,Fiebig05,Khomskii06,Cheong07,Tokura07,Kimura07}. Spiral spin orders in insulating magnets with frustrated interactions induce ferroelectric polarization $\bm P$ via an inverse effect of the Dzyaloshinskii--Moriya interaction, which, in the presence of finite vector spin chirality $\bm m_i \times \bm m_j$, causes a local electric polarization
\begin{equation}
\bm p_{ij} \propto \bm e_{ij} \times (\bm m_i \times \bm m_j),
\label{eq:KNBM}
\end{equation}
where $\bm m_i$ represents the magnetization vector on a magnetic ion, and $\bm e_{ij}$ the unit vector connecting adjacent magnetic ions,~\cite{Katsura05,Mostovoy06,Sergienko06}. The mutual coupling between magnetic and electric orders in multiferroics, the so-called magnetoelectric coupling, gives rise to rich cross-correlation phenomena.

A staggered orbital ordering in the orthorhombically distorted crystal lattice of TbMnO$_3$ causes significant antiferromagnetic interactions $J_b$ on the next-nearest-neighbor Mn-Mn bonds along the $b$ axis, which compete with the nearest-neighbor ferromagnetic interactions $J_{ab} $ [Fig.~\ref{Fig1}(a)]~\cite{Kimura03b}. This magnetic frustration stabilizes a rotating alignment of Mn spins with a propagation vector $\bm Q_m$$\parallel$$\bm b$. In the absence of an external magnetic field, the Mn spins in TbMnO$_3$ rotate within the $bc$ plane to form a transverse spiral (cycloidal) order~\cite{Kenzelmann05}, which is accompanied by a ferroelectric polarization $\bm P$$\parallel$$\bm c$ ($+\bm P_c$ or $-\bm P_c$) where the sign depends on the rotational sense of the spin spiral [Fig.~\ref{Fig1}(b)]~\cite{Yamasaki07a}. If a magnetic field $\bm H$$\parallel$$\bm Q_m$ ($+\bm H_b$ or $-\bm H_b$) is applied, a 90$^{\circ}$ flop of $\bm P$ from $\bm P$$\parallel$$\bm c$ to $\bm P$$\parallel$$\bm a$ ($+\bm P_a$ or $-\bm P_a$) appears in the experimental phase diagram of TbMnO$_3$ [see Fig.~\ref{Fig1}(c)]~\cite{Kimura05}.

This $\bm P$ flop can be attributed to $\bm H$-induced changes of the magnetic structure~\cite{Aliouane09,Yamasaki08}, but it contains several puzzles. First the emergence of $\bm P$$\parallel$$\bm a$ under $\bm H$$\parallel$$\bm b$ cannot be understood simply. If $\bm H$ is applied along $\bm Q_m$($\parallel$$\bm b$), as illustrated in Fig.~\ref{Fig1}(d), a longitudinal conical spin order with net magnetization $\bm M$($\parallel$$\bm H$) is expected to be stabilized by a gain in the Zeeman energy. This conical Mn-spin order, however, cannot induce a finite $\bm P$ via the inverse Dzyaloshinskii--Moriya mechanism exemplified by Eq.~(\ref{eq:KNBM}) because $\bm m_i \times \bm m_j$ and $\bm e_{ij}$($\parallel$$\bm b$) are parallel to each other. Instead, the $ab$-plane cycloidal spin state [see Fig.~\ref{Fig1}(e)] was proposed as a possible magnetic structure in the $\bm P$$\parallel$$\bm a$ phase of TbMnO$_3$~\cite{Aliouane09,Yamasaki08}, but its mechanism has not been fully clarified. Similar 90$^\circ$ $\bm P$ flops with $\bm H$$\parallel$$\bm Q_m$ have been reported for several other spin-spiral multiferroic materials such as DyMnO$_3$~\cite{Goto04}, LiCu$_2$O$_2$~\cite{SPark07}, MnWO$_4$~\cite{Taniguchi08}, and Ni$_3$V$_2$O$_8$~\cite{Kenzelmann06}.

Second, a memory effect in the direction of $\bm H$-induced $\bm P$ flop has been reported~\cite{Abe07,Senff08}. Naively, it is believed that the direction of $\bm P$ after the flop, i.e., $+\bm P_a$ or $-\bm P_a$, should be selected in a {\it probabilistic} way because these two states are degenerate in energy. Several experiments, however, have reported that if one increases and/or decreases a magnetic field, the $\bm P$-flop direction is seemingly selected in a {\it deterministic} way depending on the history of the $\bm H$ sweep. The deterministic nature of the flop with a unique correspondence $+\bm P_c$$\rightarrow$$+\bm P_a$ and $-\bm P_c$$\rightarrow$$-\bm P_a$ has been confirmed by real-space imaging of ferroelectric domains using second harmonic generation (SHG) microscopy~\cite{Matsubara15}. This memory effect was previously attributed to $f$--$d$ coupling between Mn spins and Tb $f$-electron moments~\cite{Abe07,Prokhnenko07a}. However, a similar memory effect was reported also for the $\bm H$-induced $\bm P$ flop in CuFeO$_2$~\cite{Mitamura07} and MnWO$_4$~\cite{Taniguchi09,Mitamura12} without $f$-electron moments, where the effect was attributed to extrinsic origins such as repopulation of multiple Q domains~\cite{Mitamura07,Mitamura12} or nuclei growth of ferroelectric embryos~\cite{Taniguchi09}.

As observations of the $\bm H$-induced $\bm P$ flop and the memory effect have been reported for several spin-spiral multiferroic materials, we expect that there is a general physical origin intrinsically underpinning them. In this paper, we theoretically study both the statics and dynamics of the $\bm H$-induced $\bm P$ flop in TbMnO$_3$ under $\bm H$$\parallel$$\bm b$ by utilizing a microscopic spin model. A Monte-Carlo calculation of this spin model reproduces the observed $\bm P$ flop from $\bm P$$\parallel$$\bm c$ to $\bm P$$\parallel$$\bm a$, and uncovers that the magnetic structure in the $\bm P$$\parallel$$\bm a$ phase is not a naively believed, simple $ab$-plane cycloid but a nearly transverse conical state.
Moreover, numerical simulations of the phase-transition dynamics reveal that the flop direction of spin-spiral plane under an external $\bm H$ field is governed by magnetic torques from $\bm H$ acting on local Mn spins, which results in a deterministic $\bm P$ flop. Our finding may enable controllable switching of $\bm P$ direction via magnetic fields towards potential applications of multiferroic materials to storage devices.

\section{Spin Model}
To describe the Mn-spin system in TbMnO$_3$, we employ a classical Heisenberg model on the orthorhombic lattice containing frustrated exchange interactions, Dzyaloshinskii--Moriya interactions, single-ion magnetic anisotropies, and Zeeman coupling~\cite{Mochizuki09a,Mochizuki09b,Mochizuki10b,Mochizuki11}. The Mn spins are treated as classical magnetization vectors $\bm m_i$(=$-\bm S_i/\hbar$) with norm of $m$=2. The full Hamiltonian consists of five terms:
\begin{eqnarray}
\mathcal{H}=\mathcal{H}_{\rm ex}+\mathcal{H}_{\rm DM}
+\mathcal{H}_{\rm sia}^D+\mathcal{H}_{\rm sia}^E+\mathcal{H}_{\rm Zeeman},
\end{eqnarray}
where
\begin{eqnarray}
\mathcal{H}_{\rm ex} &=&\sum_{<i,j>} J_{ij} \bm m_i \cdot \bm m_j,\\
\mathcal{H}_{\rm DM}&=&
\sum_{<i,j>}\bm D_{ij}\cdot(\bm m_i \times \bm m_j),\\
\mathcal{H}_{\rm sia}^D&=&D \sum_{i} m_{\zeta i}^2,\\
\mathcal{H}_{\rm sia}^E &=&
E\sum_{i}(-1)^{i_x+i_y}(m_{\xi i}^2-m_{\eta i}^2),\\
\mathcal{H}_{\rm Zeeman}&=&-g \mu_{\rm B} \bm H \cdot \sum_{i} \bm m_i.
\end{eqnarray}
Here $i_x$, $i_y$, and $i_z$ are the integer coordinates of the $i$th Mn ion with respect to the pseudocubic $x$, $y$, and $z$ axes.

The first term $\mathcal{H}_{\rm ex}$ describes the exchange interactions [see Fig.~\ref{Fig1}(f)]. The frustration between the nearest-neighbor ferromagnetic exchange $J_{ab}$ and the next-nearest-neighbor antiferromagnetic exchange $J_b$ results in transverse spiral (cycloidal) orders of Mn spins~\cite{Kimura03b,Picozzi06}, whereas the inter-plane antiferromagnetic coupling $J_c$ causes their staggered stacking. 

The second term $\mathcal{H}_{\rm DM}$ denotes the Dzyaloshinskii--Moriya interactions~\cite{Dzyaloshinskii58,Moriya60a,Moriya60b}. The Moriya vectors $\bm D_{ij}$ on the Mn-O-Mn bonds are expressed using five parameters because of the crystal symmetry~\cite{Solovyev96}; $\alpha_{ab}$, $\beta_{ab}$ and $\gamma_{ab}$ for those on the in-plane Mn-O-Mn bonds, and $\alpha_c$ and $\beta_c$ for those on the inter-plane Mn-O-Mn bonds [see Fig.~\ref{Fig1}(f)]. Their expressions are given by
\begin{eqnarray}
\bm D_{i,i+\hat x}&=&\left[
\begin{array}{c}
-(-1)^{i_x+i_y+i_z}\alpha_{ab} \\
(-1)^{i_x+i_y+i_z}\beta_{ab} \\
(-1)^{i_x+i_y}\gamma_{ab} \\
\end{array}
\right], \\
\bm D_{i,i+\hat y}&=&\left[
\begin{array}{c}
(-1)^{i_x+i_y+i_z}\alpha_{ab} \\
(-1)^{i_x+i_y+i_z}\beta_{ab} \\
(-1)^{i_x+i_y}\gamma_{ab} \\
\end{array}
\right], \\
\bm D_{i,i+\hat z}&=&\left[
\begin{array}{c}
(-1)^{i_z}\alpha_c \\
(-1)^{i_x+i_y+i_z}\beta_c \\
0 \\
\end{array}
\right].
\label{eq:DMVECS}
\end{eqnarray}

The terms $\mathcal{H}_{\rm sia}^D$ and $\mathcal{H}_{\rm sia}^E$ stand for single-ion magnetic anisotropies, defined with tilted local axes $\xi_i$, $\eta_i$ and $\zeta_i$ attached to the $i$th MnO$_6$ octahedron. $\mathcal{H}_{\rm sia}^D$ renders the magnetization along the $c$ axis hard. On the other hand, $\mathcal{H}_{\rm sia}^E$ causes an alternate arrangement of the local hard and easy magnetization axes in the $ab$ plane because of the staggered orbital ordering. Using the experimentally measured crystal parameters in Ref.~\cite{Alonso00}, the directional vectors $\bm \xi_i$, $\bm \eta_i$ and $\bm \zeta_i$ are calculated with respect to the $a$, $b$, and $c$ axes in the $Pbnm$ setting~\cite{Mochizuki09b,Mochizuki11}. Explicit formula of the directional vectors are given by,
\begin{eqnarray}
\label{eq:tilax1}
\bm {\xi}_i&=&\left[
\begin{array}{c}
a[0.25+(-1)^{i_x+i_y}(0.75-x_{{\rm O}_2})] \\
b[0.25-(-1)^{i_x+i_y}(y_{{\rm O}_2}-0.25)] \\
c(-1)^{i_x+i_y+i_z}z_{{\rm O}_2} \\
\end{array}
\right], \\
\label{eq:tilax2}
\bm {\eta}_i&=&\left[
\begin{array}{c}
a[-0.25+(-1)^{i_x+i_y}(0.75-x_{{\rm O}_2})] \\
b[0.25+(-1)^{i_x+i_y}(y_{{\rm O}_2}-0.25)] \\
-c(-1)^{i_x+i_y+i_z}z_{{\rm O}_2} \\
\end{array}
\right], \\
\label{eq:tilax3}
\bm {\zeta}_i&=&\left[
\begin{array}{c}
-a(-1)^{i_x+i_y+i_z}x_{{\rm O}_1} \\
b(-1)^{i_z}(0.5-y_{{\rm O}_1}) \\
0.25c \\
\end{array}
\right].
\end{eqnarray}
Here $x_{{\rm O}_2}$, $y_{{\rm O}_2}$ and $z_{{\rm O}_2}$ ($x_{{\rm O}_1}$ and $y_{{\rm O}_1}$) are the coordination parameters of the in-plane (out-of-plane) oxygens, and $a$, $b$ and $c$ are the lattice parameters. Their values have been experimentally determined as $a=5.29314 \AA$, $b=5.8384 \AA$, $c=7.4025 \AA$, $x_{{\rm O}_1}=0.1038$, $y_{{\rm O}_1}=0.4667$, $x_{{\rm O}_2}=0.7039$, $y_{{\rm O}_2}=0.3262$, and $z_{{\rm O}_2}=0.0510$~\cite{Alonso00}. The last term $\mathcal{H}_{\rm Zeeman}$ represents the Zeeman coupling with an external magnetic field $\bm H$.

\begin{figure}
\includegraphics[width=1.0\columnwidth]{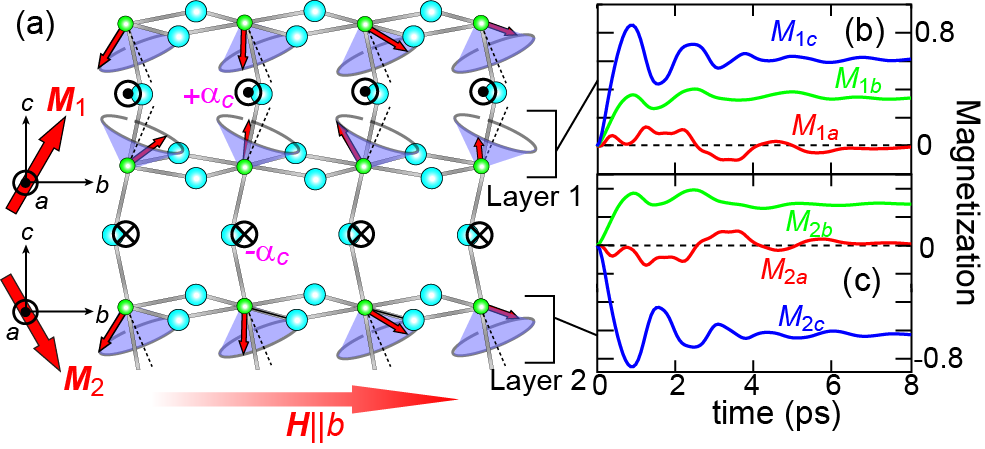}
\caption{(color online). (a) Conical magnetic structure in the $\bm P$$\parallel$$\bm a$ phase under $\bm H$$\parallel$$\bm b$ revealed by the Monte-Carlo calculation. Staggered pattern of the $a$-axis components of the Moriya vectors on the inter-plane Mn-O-Mn bonds as well as averaged magnetizations $\bm M_1$ and $\bm M_2$ in two different MnO layers (Layers 1 and 2) are also shown. (b) and (c) Simulated temporal evolutions of averaged magnetizations in (b) Layer 1 and (c) Layer 2 for the $\bm H$-induced $\bm P$ flop after application of $+\bm H_b$ of 17 T.}
\label{Fig2}
\end{figure}
The model parameters for TbMnO$_3$ were calculated microscopically in Ref.~\cite{Mochizuki09b}; ($J_{ab}$, $J_b$, $J_c$)=($-$0.74, 0.64, 1.0), ($D$, $E$)=(0.2, 0.25), ($\alpha_{ab}$, $\beta_{ab}$, $\gamma_{ab}$)=(0.1, 0.1, 0.14) and ($\alpha_c$, $\beta_c$)=(0.48, 0.1) in units of meV. The spin model with this parameter set has turned out to describe magnetoelectric phenomena in TbMnO$_3$ quantitatively. A Monte-Carlo analysis of the spin model reproduces the $bc$-plane cycloidal spin order propagating along the $b$ axis with a wave number $Q_m$=0.3$\pi$ at $\bm H$=0, which coincides with the observed $bc$-plane spin cycloid in TbMnO$_3$ ($Q_m$=0.28$\pi$)~\cite{Kenzelmann05}. The optical absorption spectra associated with electromagnon excitations for TbMnO$_3$ have been perfectly reproduced with respect to light-polarization selection rule, spectral structure with double peaks, and resonance frequencies~\cite{Mochizuki10d}. The Monte-Carlo analysis has also reproduced the phase diagrams of TbMnO$_3$ in plane of temperature and magnetic field as well as the $\bm H$-induced $\bm P$ flop from $\bm P$$\parallel$$\bm c$ to $\bm P$$\parallel$$\bm a$ under $\bm H$$\parallel$$\bm b$~\cite{Mochizuki10c}. These quantitative reproductions of experimental results guarantee the validities of the spin model and the parameter set used in the calculations and, hence, the validity of the obtained results.

\section{Results and Discussions}
In the Monte-Carlo calculation, we find that a magnetic structure in the $\bm P$$\parallel$$\bm a$ phase is not a simple $ab$-plane cycloid as commonly believed but a conical structure [Fig.~\ref{Fig2}(a)]. The cones in this state are slightly slanted from the $c$ axis towards the $\bm H$ direction ($\parallel$$\bm b$). The orthorhombic TbMnO$_3$ crystal has two different MnO layers (layers 1 and 2) stacking alternately along the $c$ axis. The averaged magnetizations $\bm M_1$ and $\bm M_2$ for layers 1 and 2 are calculated from $\bm M_\gamma=(1/N_xN_y)\sum^{\prime}_i \bm m_{i\gamma}$, where the summation $\sum^{\prime}_i$ is taken over the Mn sites within the $\gamma$th layer. The orientations of $\bm M_1$ and $\bm M_2$ are depicted by thick arrows in Fig.~\ref{Fig2}(a). Their $c$-axis components are staggered, whereas their $b$-axis components uniformly point in the $\bm H$ direction. This spin structure results from a compromise of several competing interactions. The rotating spin alignment is caused by the frustrated exchange interactions $\mathcal{H}_{\rm ex}$, whereas the spin rotation residing nearly within the $ab$ plane is favored by the easy-plane magnetic anisotropy $\mathcal{H}_{\rm sia}^D$. The uniform $b$-axis components $M_{\gamma b}$ result from the Zeeman coupling $\mathcal{H}_{\rm Zeeman}$ with $\bm H$. The staggered $c$-axis components $M_{\gamma c}$ are stabilized by the Dzyaloshinskii--Moriya interactions $\mathcal{H}_{\rm DM}$ in the presence of the uniform $M_{\gamma b}$ because the Moriya vectors on the inter-plane Mn-O-Mn bonds have large $a$-axis components with alternate signs, $+\alpha_c$ or $-\alpha_c$ [see Fig.~\ref{Fig2}(c)].

Peculiarity of this magnetic structure will show up in optical spectra. Electromagnon excitations in multiferroic Mn perovskites with spiral spin orders such as TbMnO$_3$, DyMnO$_3$, and Eu$_{1-x}$Y$_x$MnO$_3$ are ascribed to the symmetric exchange striction mechanism due to coupling between the ac electric field $\bm E^\omega$ of light and local electric polarizations $\bm p_{ij}=\bm \Pi_{ij} \bm m_i \cdot \bm m_j$ on the distorted Mn-O-Mn bonds~\cite{Aguilar09,Mochizuki10a}. (Note that these exchange-striction driven local polarizations are different from those given by Eq.~(1), which cancel out in total and never contribute to the ferroelectricity.) They can be activated when $\bm E^\omega$ is parallel to the $a$ axis irrespective of the spiral-plane orientation as long as cycloidal spin orders in $R$MnO$_3$ are considered. In Ref.~\cite{Rovillain11}, electromagnon spectra for the $\bm P$$\parallel$$\bm a$ phase of TbMnO$_3$ under $\bm H$$\parallel$$\bm b$ was investigated theoretically. In this study, although the detailed magnetic structure was not explicitly argued, we expect the $ab$-plane conical phase because the spin model and the parameter set used in Ref.~\cite{Rovillain11} are totally the same as those used in the present study. It was confirmed that the light-polarization selection rule holds even for the $\bm P$$\parallel$$\bm a$ phase with the $ab$-plane conical order, that is, the electromagnons are activated only with $\bm E^\omega$$\parallel$$\bm a$ via the symmetric exchange striction. Moreover, it was predicted that the higher-energy spectral peak splits into two peaks. This prediction was indeed confirmed by the Raman spectroscopy~\cite{Rovillain11}. The coincidence between the theoretical prediction and the experimental observation strongly supports the validity of the predicted $ab$-plane conical spin structure because splitting of the higher-lying electromagnon spectral peak can be expected only for the $ab$-plane conical state but not for the $ab$-plane cycloidal state~\cite{Mochizuki10a}.

Next, we investigate the dynamics of $\bm m_i$ and $\bm p_{ij}$ for the $\bm H$-induced $\bm P$ flop in TbMnO$_3$ using the Landau--Lifshitz--Gilbert equation given by
\begin{equation}
\frac{d \bm m_i}{d t}=-\bm m_i \times \bm H^{\rm eff}_i
+ \frac{\alpha_{\rm G}}{m} \bm m_i \times \frac{d \bm m_i}{d t}.
\label{eq:LLGEQ}
\end{equation} 
Here $\alpha_{\rm G}$ is the dimensionless Gilbert-damping coefficient, which is fixed at 0.05 in the calculations~\cite{Mochizuki10a}. The effective local magnetic field $\bm H^{\rm eff}_i$ acting on the $i$th Mn magnetization $\bm m_i$ is derived from the Hamiltonian $\mathcal{H}$ as
\begin{equation}
\bm H^{\rm eff}_i = - \partial \mathcal{H} / \partial \bm m_i.
\label{eq:EFFMF}
\end{equation}
We solve this equation numerically using the fourth-order Runge-Kutta method with a system of 40$\times$40$\times$36 sites on which periodic boundary conditions are imposed. Spatiotemporal distributions of local polarizations $\bm p_{ij}$ are calculated from simulated configurations of $\bm m_i$ using Eq.~(\ref{eq:KNBM}) by considering the inverse Dzyaloshinskii--Moriya mechanism as their origin~\cite{Malash08}.

\begin{figure*}
\includegraphics[width=2.0\columnwidth]{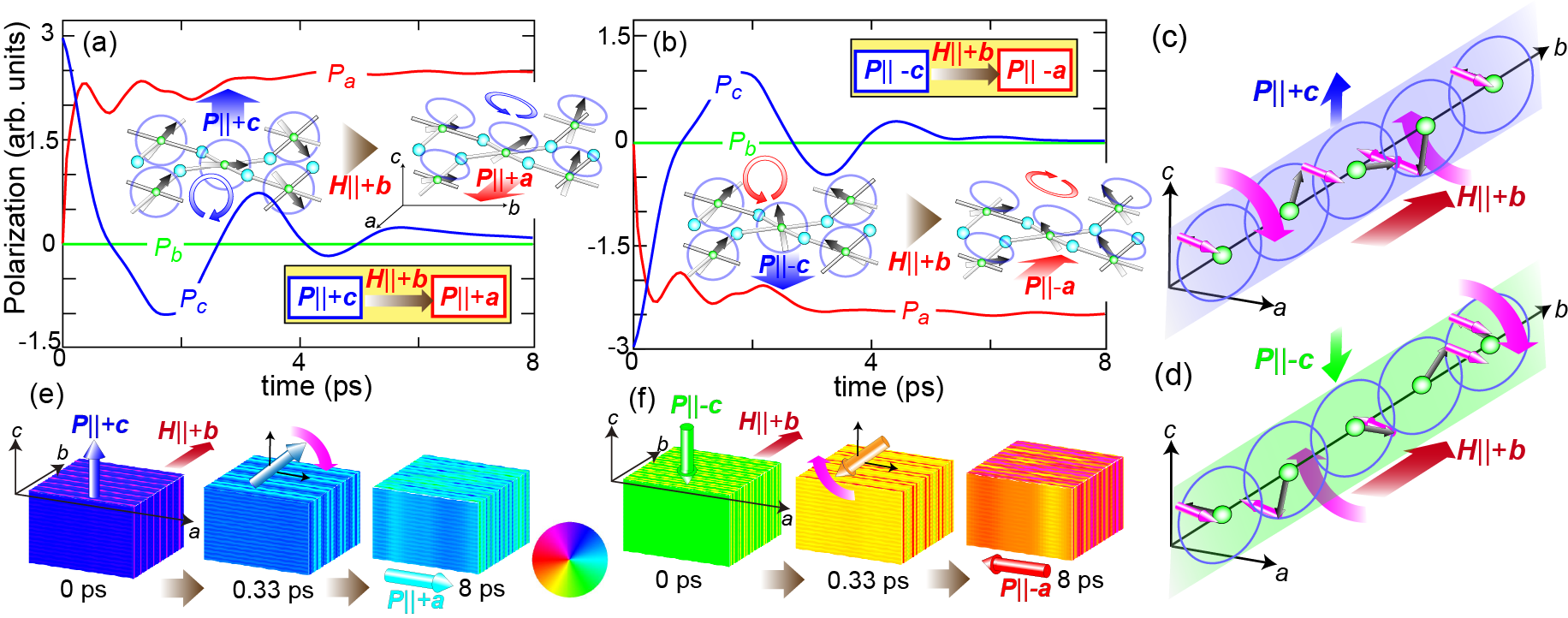}
\caption{(color online). (a), (b) Simulated temporal evolution of ferroelectric polarization $\bm P=(P_a, P_b, P_c)$ for the $\bm H$-induced $\bm P$ flop after $+\bm H_b$ of 17 T is applied. $+\bm P_c$ necessarily flops to $+\bm P_a$ (a), whereas $-\bm P_c$ flops to $-\bm P_a$ (b) in a deterministic way. (c), (d) Directions of magnetic torques $\bm \tau_i$$\propto$$-\bm m_i \times \bm H$ acting on local Mn magnetizations under $+\bm H_b$ are represented by solid arrows pointing in the $+a$ and $-a$ directions for (c) clockwise and (d) counterclockwise $bc$-plane spin cycloids. The magnetic torques cooperatively contribute to rotate the cycloidal plane in a counterclockwise fashion around the $b$ axis irrespective of the spin-cycloidal sense. (e), (f) Snapshots of simulated spatiotemporal dynamics of local electric polarizations $\bm p_{\ij}$ for the $\bm P$ flop (e) from $+\bm P_c$ to $+\bm P_a$ and (f) from $-\bm P_c$ to $-\bm P_a$.}
\label{Fig3}
\end{figure*}
Repeating the simulation with different initial configurations of $\bm m_i$, we find that the flop occurs in a deterministic way. In Fig.~\ref{Fig3}(a) and (b), we show simulated time profiles of the $a$-, $b$-, and $c$-axis components of ferroelectric polarization $\bm P(t)=(1/N)\sum_i \bm p_i(t)$ for the $\bm H$-induced $\bm P$ flop when $+\bm H_b$ of 17 T is applied. It turns out that $+\bm P_c$ ($-\bm P_c$) always flops to $+\bm P_a$ ($-\bm P_a$) under $+\bm H_b$, in contradistinction to the naively anticipated probabilistic flop.

This deterministic $\bm P$ flop can be understood as follows: if $\bm H$ is applied, each Mn magnetization $\bm m_i$ experiences a magnetic torque $\bm \tau_i$$\propto$$-\bm m_i \times \bm H$. For the $bc$-plane spin cycloids with $+\bm P_c$ and $-\bm P_c$, the directions of the torques under $+\bm H_b$ are marked by solid arrows pointing in the $+a$ and $-a$ directions in Fig.~\ref{Fig3}(c) and (d), respectively. For both cases, the torques cooperatively contribute to rotate the cycloidal plane in a counterclockwise fashion around the $b$ axis, resulting in the deterministic switching from $+\bm P_c$ to $+\bm P_a$ or from $-\bm P_c$ to $-\bm P_a$. Figure~\ref{Fig3}(e) shows simulated spatiotemporal dynamics of local $\bm p_{ij}$ vectors for the switching from $+\bm P_c$ to $+\bm P_a$, whereas Fig.~\ref{Fig3}(f) shows that from $-\bm P_c$ to $-\bm P_a$ under $+\bm H_b$ of 17 T. 

\begin{figure}[h]
\includegraphics[width=1.0\columnwidth]{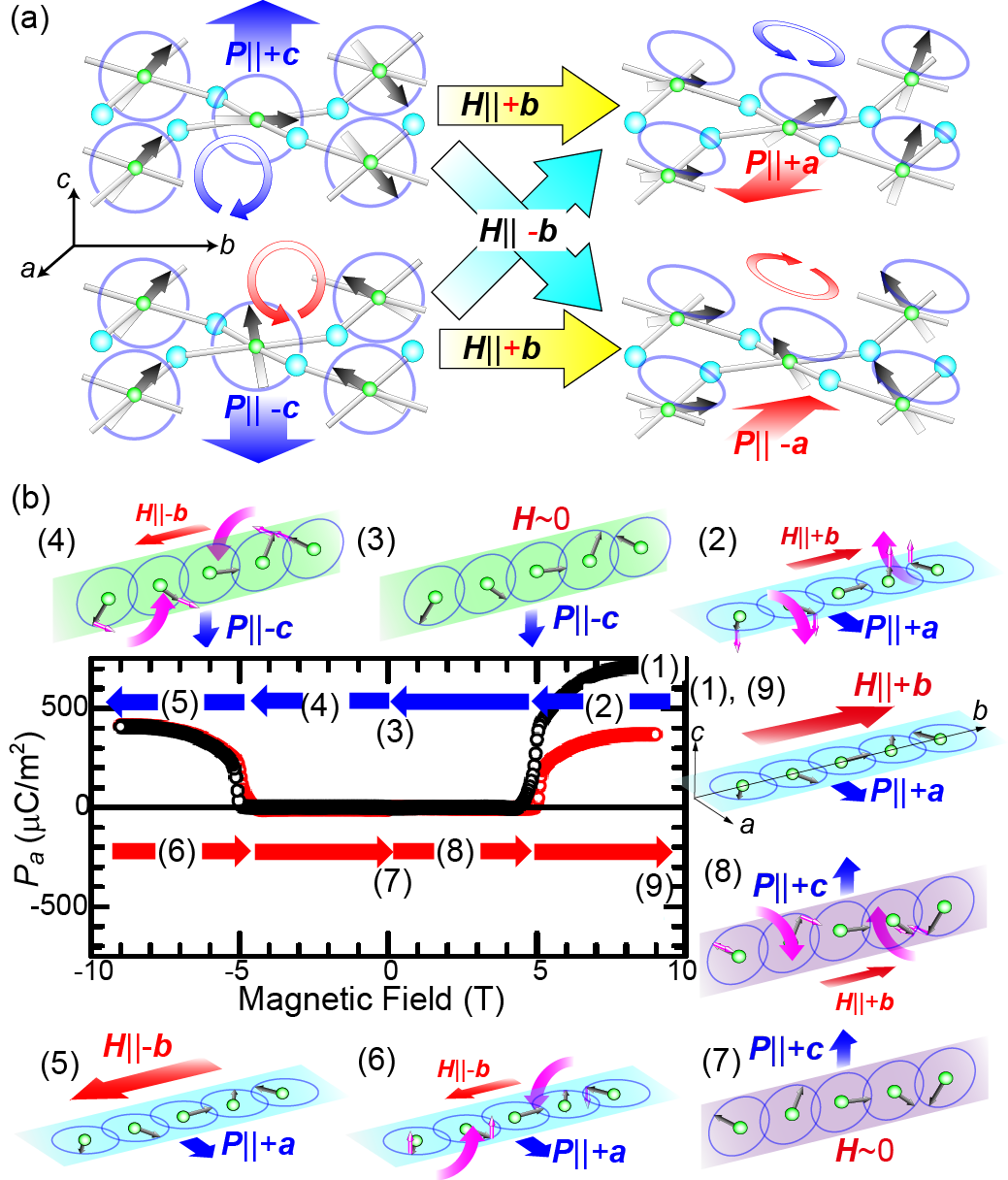}
\caption{(color online). (a) Deterministic $\bm H$-induced $\bm P$ flops in TbMnO$_3$. When $+\bm H_b$ is applied, $+\bm P_c$ ($-\bm P_c$) necessarily flops to $+\bm P_a$ ($-\bm P_a$). If the sign of $\bm H$ is changed, these relationships become reversed, i.e., $+\bm P_c$ ($-\bm P_c$) necessarily flops to $-\bm P_a$ ($+\bm P_a$). (b) Experimental profile of $P_a$ when one decreases $H_b$ from $+9$ T to $-9$ T and then increases $H_b$ from $-9$ T to $+9$ T~\cite{Abe07}. Expected spin-cycloidal planes, orientations of $\bm P$ and $\bm H$, directions of magnetic torques, and rotating directions of the cycloidal plane are indicated for each stage of the magnetic-field sweep. Here, magnetic structures in the $\bm P$$\parallel$$\bm a$ phases are presented as $ab$-plane spin cycloids for simplicity, although significant conical modulations exist.}
\label{Fig4}
\end{figure}
We also find that if $\bm H$ is reversed, the relationship is reversed, as indicated in Fig.~\ref{Fig4}(a); $+\bm P_c$ flops to $-\bm P_a$, whereas $-\bm P_c$ flops to $+\bm P_a$ under $-\bm H_b$. This occurs because the directions of the magnetic torques are determined by the sign of $\bm H$ and become opposite if the sign of $\bm H$ is reversed. Subsequently, the spin-spiral plane rotates in a clockwise fashion under $-\bm H_b$, again, irrespective of the spin-spiral sense.

The experimentally observed memory effect in the $\bm H$-induced $\bm P$ flop in TbMnO$_3$ can be explained by the above mechanism. Figure~\ref{Fig4} shows the measured evolution of the $a$-axis component of $\bm P$ ($P_a$) during the sweep of the magnetic field $\bm H$$\parallel$$\bm b$, previously reported in Ref.~\cite{Abe07}. Starting from $+\bm H_b$ of $+9$ T, Abe $et$ $al$ gradually decreased $H_b$ to $-9$ T and then increased $H_b$ from $-9$ T to $+9$ T. The $\bm P$$\parallel$$\bm a$ phase appears in the range of $H_b>5$ T and $H_b<5$ T. They found that the $P_a$ always has positive values if they start from the $+\bm P_a$ state. This behavior is a consequence of the clockwise rotation of the spiral plane under $+\bm H_b$ at stages (2) and (8), and the counterclockwise rotation under $-\bm H_b$ at stages (4) and (6), which can be experimentally confirmed by simultaneous measurement of the $c$-axis component $P_c$.

Here it is noteworthy that the memory effect on the $\bm P$ direction can be observed not only as a function of $\bm H$ field but also as a function of temperature as reported for thermally induced ferroelectric transitions in MnWO$_4$~\cite{Meier09} and CuO~\cite{WuWB10} at $\bm H$=0. Since the magnetic torque from external $\bm H$ field is absent in these cases, the observed effect can be attributed to other mechanisms such as nuclei growth of tiny domains survived due to pinning. This extrinsic mechanism can work also in the $\bm H$-induced $\bm P$ flop even while the magnetic torque is a major source of the memory effect. We expect that comprehensive understanding of memory effects in multiferroics can be achieved by taking both intrinsic and extrinsic mechanisms into account.

The deterministic $\bm P$ flop in TbMnO$_3$ results in the emergence of nominally charged domain walls as observed by the SHG microscopy~\cite{Matsubara15}. At $\bm H$=0, multiferroic domain walls between $+\bm P_c$ and $-\bm P_c$ domains in the $ac$ plane tend to run parallel to the $c$ axis, because of the strong inter-plane antiferromagnetic coupling $J_c$ along the $c$ axis. Under this condition, 180$^{\circ}$ domain walls between $+\bm P_c$ and $-\bm P_c$ at $\bm H$=0 necessarily turn into head-to-head or tail-to-tail domain walls between $+\bm P_a$ and $-\bm P_a$ upon the $P$-flop driven by $\bm H$$\parallel$$\bm b$. Across the domain wall, we expect that the spin-spiral plane continuously changes from $+ab$ (clockwise) to $-ab$ (anticlockwise) via $bc$~\cite{Kagawa09,Leo15}. Namely the $bc$-plane spin cycloid appears between the two different $ab$-plane cycloids. Consequently, the electric polarization gradually varies from $+\bm P_a$ to $-\bm P_a$ via $\bm P$$\parallel$$\bm c$ to form a Neel wall of $\bm p_{ij}$. Thus, the domain wall between $+\bm P_a$ and $-\bm P_a$ states can be regarded as head-to-head or tail-to-tail only in the macroscopic scale, and there is no {\it classical} divergence of the electrostatic potential at the center of the wall. However, we can still expect small but finite $-\bm \nabla \cdot \bm P$ between sites in the domain-wall area owing to the discreteness of the crystal lattice, which causes nominally charged multiferroic domain walls.

\section{Summary}
In summary, we theoretically studied the magnetic-field-induced 90$^\circ$ flop of ferroelectric polarization $\bm P$ in the spin-spiral-based multiferroic material TbMnO$_3$. Our Monte-Carlo analysis of a microscopic spin model revealed that the magnetic structure in the field-induced $\bm P$$\parallel$$\bm a$ phase is not the simple $ab$-plane spin cycloid as commonly believed, but a conical spin structure, which is stabilized by the Zeeman coupling and the Dzyaloshinskii--Moriya interactions associated with vectors on the inter-plane Mn-O-Mn bonds. Our simulations of the flop dynamics indicated that the flop direction of $\bm P$ or the rotation direction of the spiral plane is governed by the sign of the applied magnetic field via the magnetic torques acting on local Mn magnetizations. Consequently, the choice of $+\bm P_a$ or $-\bm P_a$ after the flop is selected in a deterministic way depending on the magnetic-field direction. This contrasts the naively believed probabilistic flop and accounts for the reported memory effect; that is, the $\bm P$ direction depends on the history of the magnetic-field sweep. This finding enables controllable switching of the $\bm P$ direction by application of magnetic fields and paves the way to technical application of spin-spiral multiferroics.

This research was in part supported by JSPS KAKENHI (Grant Nos. 25870169 and 25287088). The author thanks D. Meier, M. Matsubara, T. Kimura, S. Manz, and M. Fiebig for fruitful discussions.

\end{document}